\documentclass[oldversion]{aa}
\usepackage{graphicx}
\usepackage{txfonts}
\usepackage{natbib}
\bibpunct{(}{)}{;}{a}{}{,} 

\usepackage{graphicx}
\usepackage{siunitx}

\usepackage{natbib,twoopt}
\usepackage[breaklinks=true]{hyperref} 
\bibpunct{(}{)}{;}{a}{}{,} 
\makeatletter
\newcommandtwoopt{\citeads}[3][][]{\href{http://adsabs.harvard.edu/abs/#3}%
{\def\hyper@linkstart##1##2{}%
\let\hyper@linkend\@empty\citealp[#1][#2]{#3}}}
\newcommandtwoopt{\citepads}[3][][]{\href{http://adsabs.harvard.edu/abs/#3}%
{\def\hyper@linkstart##1##2{}%
\let\hyper@linkend\@empty\citep[#1][#2]{#3}}}
\newcommandtwoopt{\citetads}[3][][]{\href{http://adsabs.harvard.edu/abs/#3}%
{\def\hyper@linkstart##1##2{}%
\let\hyper@linkend\@empty\citet[#1][#2]{#3}}}
\newcommandtwoopt{\citeyearads}[3][][]%
{\href{http://adsabs.harvard.edu/abs/#3}
{\def\hyper@linkstart##1##2{}%
\let\hyper@linkend\@empty\citeyear[#1][#2]{#3}}}
\makeatother
\begin{document}

  \title{Keeping up with the cool stars: One TESS year in the life of AB~Doradus}

  \author{P. Ioannidis
     \and
     J.H.M.M. Schmitt }

  \institute{Hamburger Sternwarte Universit\"at Hamburg, Gojenbergsweg 112, 21029 Hamburg, Germany\\
       \email{pioannidis@hs.uni-hamburg.de}  
       }

  \date{Received 15/04/2020; accepted 20/10/2020}

 
 \abstract
{The long-term, high precision photometry delivered by the Transiting Exoplanet Survey 
Satellite (TESS) enables us to gain new insight into known and hitherto well-studied stars. In this paper,
we present the result of our TESS study of the photospheric activity of the rapid rotator AB~Doradus. 
Due to its favorable position near the southern ecliptic pole, the TESS satellite recorded almost 
600 rotations of AB~Doradus with high cadence, allowing us to study starspots and flares on this
ultra-active star. The observed peak-to-peak variation of the rotational modulations reaches almost 11\%, 
and we find that the starspots on AB~Doradus show highly preferred longitudinal positions. Using spot modeling, we measured the positions of the active regions on AB~Doradus 
and we find that preferred spot configurations should include large regions extending from low to high stellar latitudes. We interpret the apparent movement of spots as the result of both differential rotation and spot evolution and argue that the typical spot lifetimes should range between 10 and 20 days. 
We further find a connection between the flare occurrence on AB~Doradus and the visibility of the active regions on its surface, and we finally recalculated the star's rotation period using different methods and we compared it with previous determinations. 
}


  \maketitle
%

\section{Introduction}

While the study of stellar activity has been the focus of many 
observational and theoretical research efforts over the last decades, 
we are still not in a position to fully understand and cope with all of its effects 
on time-series observations. The latter has been revolutionized during
the last decade as a consequence of the availability of high-precision,
long term, and essentially uninterrupted 
space-based photometry, as delivered by the CoRoT \citep{auvergne_2009}, {\it Kepler} \citep{borucki_2010},
and Transiting Exoplanet Survey 
Satellite (TESS) missions \citep{ricker_2015}, which provided and still provides data of unsurpassed quality that is unachievable from the ground.

In such photometric observations, one typically sees light curve modulations
caused by the periodic appearance and disappearance of active regions in the
stars' photosphere, that is to say star spots. As we know from the Sun, 
sunspots rotate with the Sun and produce depressions in the observed solar flux (see an overview by \cite{froehlich_2012}). Therefore, one expects
stellar observations to provide similar information on star spots, and
many studies are dedicated to understanding how and where star spots form, how long they live, and
how those characteristics depend on the age and type of the underlying star. 

A great new research opportunity has opened up with TESS \citep{ricker_2015}.
With its anticipated observation strategy, TESS will cover $>$80\% of the sky and
therefore, unlike the cases of CoRoT and {\it Kepler}, 
also deliver high-precision space-based photometry for brighter stars that are
already well known and intensely studied at other wavelengths.
One such star is AB~Doradus, which is a bright, nearby, rapidly rotating, and very active star. It is fortuitously located at rather high ecliptic latitudes and
was almost continuously observed by TESS in its first year of operation.
In this paper, we present almost one year (298 days)
of TESS observations of \object{AB~Dor} and carry out an in-depth analysis of its photometric
activity features, as observed by TESS.


\begin{table}
\caption{Basic information on AB~Dor.}       
\label{tab:ABDor}   
\centering             
\begin{tabular}{l r}    
\hline\hline         
  R.A. & \ang{05;28;44.9} \\   
  Dec & \ang{-65;26;55.3} \\
  Gaia (DR2) & 4660766641264343680 \\
  TIC & 149248196 \\
  m$_v$ & 7 mag \\ 
\hline                  
\end{tabular}
\end{table}

\section{Observations}

\subsection{The TESS mission}

The TESS mission is dedicated to the discovery of transiting exoplanets around bright stars \citep[see][]{ricker_2015} via photometry. To 
this end, TESS was designed to cover a very large sky area (\ang{24} x \ang{96}) utilizing four
wide-field cameras. The initial two-year observation strategy of TESS includes 26
sectors, 13 of which are in the southern ecliptic hemisphere, the other 13 are in the northern ecliptic
hemisphere, and each is continuously observed for 27 days. As a result,
those parts of the sky that are closer to the ecliptic poles are observed for longer 
periods than 27 days with a maximum observation time of one year around the ecliptic
poles.
 TESS's primary scientific goal is to survey the brightest 200,000 stars for transiting exoplanets.
The actual light curves of those bright stars are produced by the Science Processing 
Operation Center (SPOC) pipeline \citep[see,][]{Jenkins2016} and are publicly accessible and available for
download from the Mikulski Archive for Space Telescopes (MAST)\footnote{https://mast.stsci.edu}

 \begin{figure}
  \centering
  \includegraphics[width=\hsize]{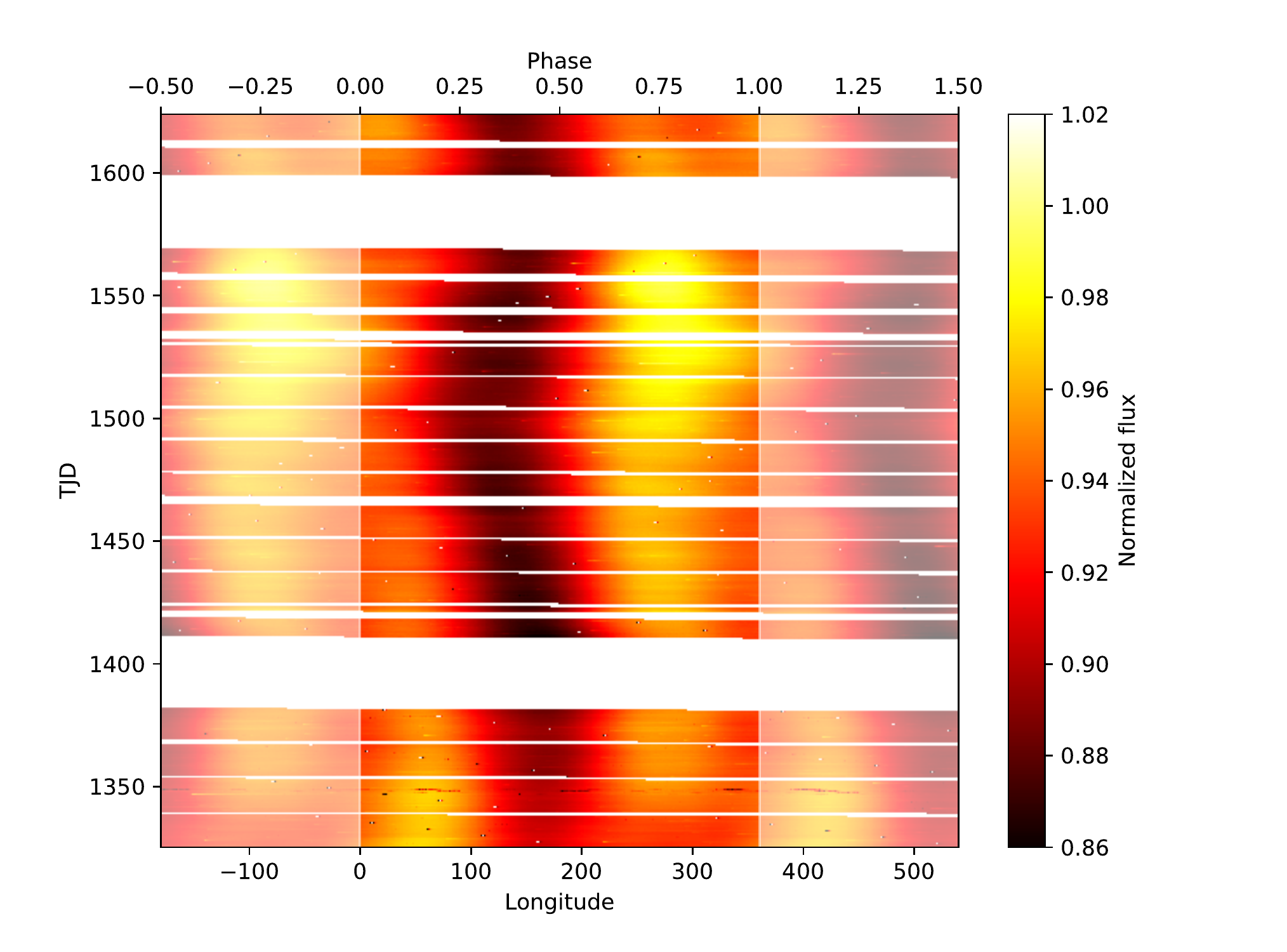}
   \caption{Total TESS time series of AB~Dor phased assuming a rotation period of $P_\mathrm{rot} = 0.514275$ days; see text for details. }
     \label{fig:0514275br}
  \end{figure}

\subsection{AB~Doradus}
\label{sect:abdor}
The star we focus on in this study is AB~Dor~A. It is the brightest member of 
a system with four stars in total \citep{guirado_2011}. However, due to the 
brightness contrast of AB~Dor A and its companions,
we ignore those in the following and thus refer to AB~Dor~A as AB~Dor for simplicity.

As a foreground star of the Large Magellanic Cloud, AB~Dor was first recognized as an interesting (flaring) X-ray stellar
source by \cite{pakull_1981}, who measured its rapid rotation (P$_{rot}$ = 0.514~d) and interpreted AB~Dor
as a new RS~CVn system. \cite{innis_1988b} presented high-resolution spectroscopy of AB~Dor, showing it
to be a single star, and the photometry of AB~Dor, covering the period from 1978 - 1987.\ It was 
suggested that ``AB~Dor had two large, long-lived starspots during this period'' \citep{innis_1988a}. 

Given its short rotation period of a little over half a day, AB~Dor can be Doppler imaged much more easily than
longer period stars.
Using observations taken in February 1989, \cite{Kuerster1994} presented the spatial distribution of starspots using Doppler images, which
also show similarities with the spot distribution discussed by \cite{innis_1988a}.\ Moreover,  \cite{Kuerster1994} showed a consistency of their pseudo-simultaneous photometry with the derived Doppler images.
Furthermore, their Doppler images suggest an inclination angle of about 60$^\circ$ between the stellar rotation axis and the line of sight. 
\cite{donati_1997} derived the differential rotation of AB~Dor and showed that it is rotating, almost as a rigid body.
Further studies of the differential rotation of AB~Dor based on Doppler imaging have been presented by
\cite{cameron_2002}, \cite{donati2003}, and \cite{jeffers_2007}.  The latter authors present a sequence of Doppler images of AB~Dor, which were constructed from data taken in December 1988, January and December 1992, and November 1993 and 1994, with special emphasis on the differential rotation of AB~Dor.

Extensive X-ray studies of AB~Dor have been presented, for example,
by \cite{kuerster_1997}, who discuss the ROSAT data on AB~Dor, and by \cite{lalitha_2013}, who present
simultaneous XMM-Newton and VLT observations of AB~Dor which, in particular, cover a very large flare. This flare was
also studied by \cite{schmitt2019}, who show the superflare nature of such flares observed by TESS during the
first two months of data taking.  
Thus, in summary, according to our current understanding, AB~Dor is rather
nearby ($d = 15.3$ pc), ultra-active, and a
pre-main-sequence object with a radius of 0.96 $\pm$ 0.06 R$_{\odot}$ \citep[see][]{guirado_2011}, which shows strong signatures of magnetic activity at all wavelengths.

\begin{table}
\caption{AB~Dor rotation periods measured by other authors using photometry and this work. The number of rotations from the maximum (unspotted) part of the light curve observed by \cite{kuerster_1997} and an equivalent one in the light curve from TESS with epoch JD = 2458473.0 (see Sect.~\ref{sect:rot} for details) are listed in the third column. }       
\label{tab:per}   
\centering             
\begin{tabular}{l c l}    
\hline\hline         
Reference & Period (days) & Rotations\\
\hline            
\cite{pakull_1981} & 0.51423 & 21184.04\\
 \cite{innis_1988a} & 0.51479 & 21161.00\\ 
 \cite{jaervinen_2005} & 0.51487 & 21157.70\\
 \cite{schmitt2019} & 0.51428 & 21182.19\\
 This work (GLS) & 0.51484 & 21159.94\\
 This work (Maxima) & 0.51429 & 21181.39\\
 This work (Minima) & 0.51423 & 21184.16\\
\hline                  
\end{tabular}
\end{table}

\section{Results}

\subsection{Light curve}

The apparent brightness of AB~Dor (m$_V$ = 7 mag) makes it one of the brightest stars observable by TESS, which, in combination
with the scintillation-free environment of space, results in a light curve with a very high signal-to-noise ratio. 
Due to the long TESS observation time of AB~Dor and its short rotation period, the traditional representation of the light curve as a time series does not allow us to explore the variations in the time series  in detail. 
Thus we chose to use a time-brightness diagram to show the total light curve of AB~Dor in a more compact and readable fashion.

For the construction of the time-brightness plot, we split the light curve in chunks of length equal to the period derived by \citet{schmitt2019}, that is, 0.514275 days. Assuming that this signal's period is indeed the star's rotation period, each of these chunks then represents one rotation of AB~Dor. Each chunk is plotted in time order from bottom to top, which provides a better impression of the light curve's temporal evolution. The bright parts of the plot show parts of the light curve with high stellar flux and, conversely, the darker parts of the plot are those parts of the light curve 
where the stellar flux is reduced. 

In Figure~\ref{fig:0514275br}, we show the time-brightness diagram constructed using the TESS data set of AB~Dor. The TESS light curve shows large peak-to-peak variations, which vary between $\sim$6\% and $\sim$11\%. We interpret those modulations as the result of star spots, which become visible as the star rotates; thus, we assume that the period of these variations is indeed the stellar rotation period.
This signal is clearly periodic, and although it almost repeats itself perfectly from one period to the next, there are evident long-term variations that alter the apparent shape of the light curve with time. We also note that Fig.~\ref{fig:0514275br} shows only the contrast between the different areas of the stellar surface, that is, 
the bright areas of the plot are not necessarily free of star spots. 

In addition to the rotation modulations, there are events with a rapid increase in the stellar flux followed
by a more or less exponential decay, which we interpret as stellar flares. 
\cite{schmitt2019} performed an in-detail analysis of those events for the first two sectors of the TESS data on AB~Dor.

 \begin{figure}
  \centering
  \includegraphics[width=\hsize]{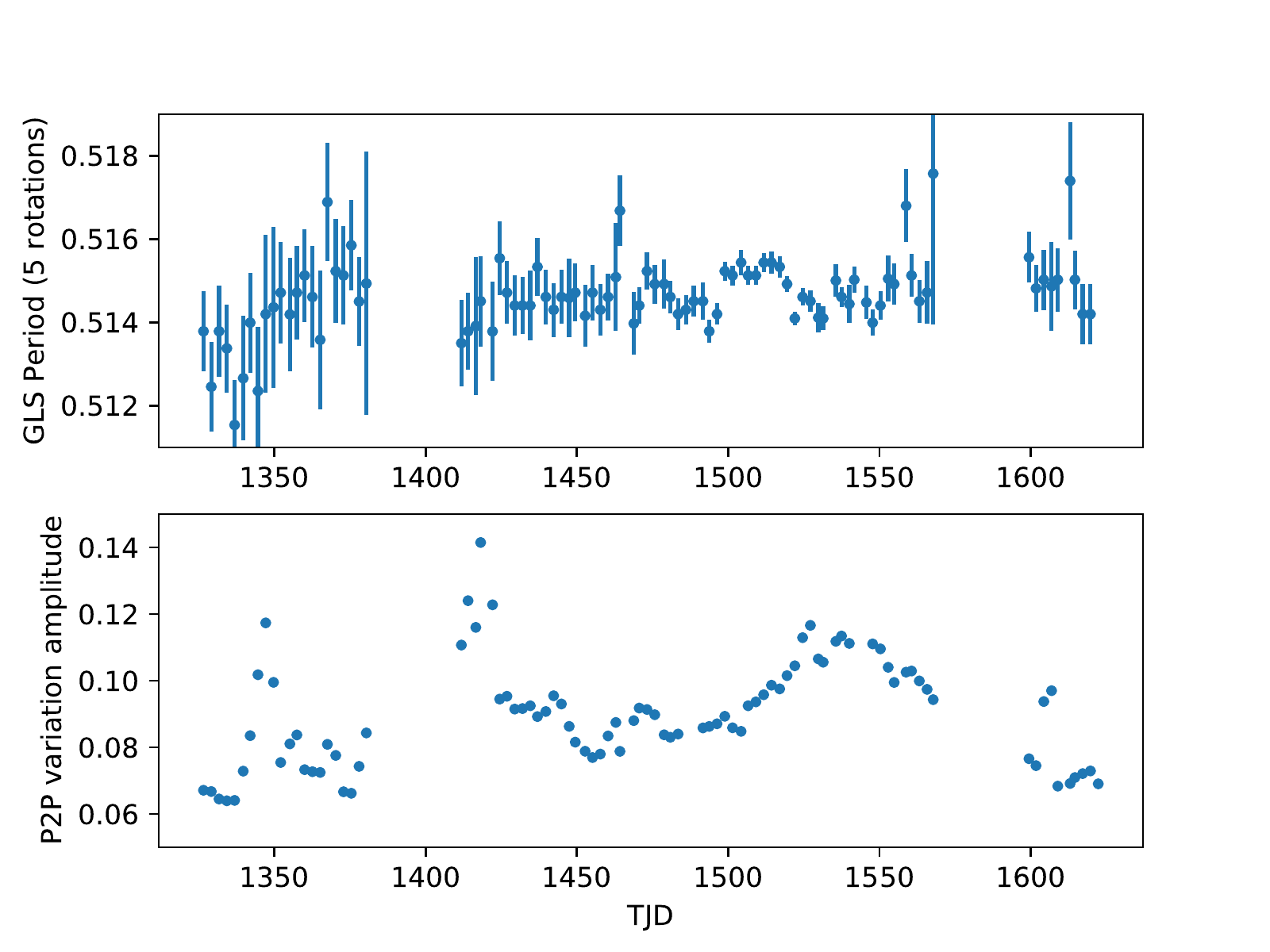}
   \caption{Top panel: Variation of the GLS periodogram of the light curve of AB~Dor, split into parts containing five rotations each. We note that the reduced error sizes in the time range between 1500$\lesssim$TJD$\lesssim$1550 are related to the shape of the light curve during that time (which is closer to sinusoidal than the rest of the light curve) and the GLS periodogram's ability to measure the periods of sinusoidal-like signals more accurately. Bottom panel: Mean peak-to-peak variation amplitude assuming sets of five rotations (the y-axis is in normalized flux units). }
     \label{fig:p2pgls}
  \end{figure}

\subsection{Activity modulations: Period and amplitude}

One of the most remarkable properties of AB~Dor is its extremely rapid rotation. Consequently, it is not surprising that its period was measured several times by different authors; Table~\ref{tab:per} provides
some of those results and their references. Given the temporal evolution of the light curve and the divergence between the different measurements through the years, we did not attempt to measure the rotation period of the star using the total light curve. Instead, we split the light curve into different parts, assuming that in each part, the star has completed five rotations, and we used the generalized Lomb-Scargle periodogram \citep[GLS; see][]{Zechmeister2009} to measure the period of the activity modulations in each one of those parts. The results of our measurements are shown in the upper panel of Fig.~\ref{fig:p2pgls}. The modulation period appears to be relatively stable during the observations with an average of $\sim$0.514843 days, except for the first $\sim$50 days; we estimate the typical uncertainty of those period
measurements to be on the order of one minute.

 \begin{figure}
  \centering
  \includegraphics[width=\hsize]{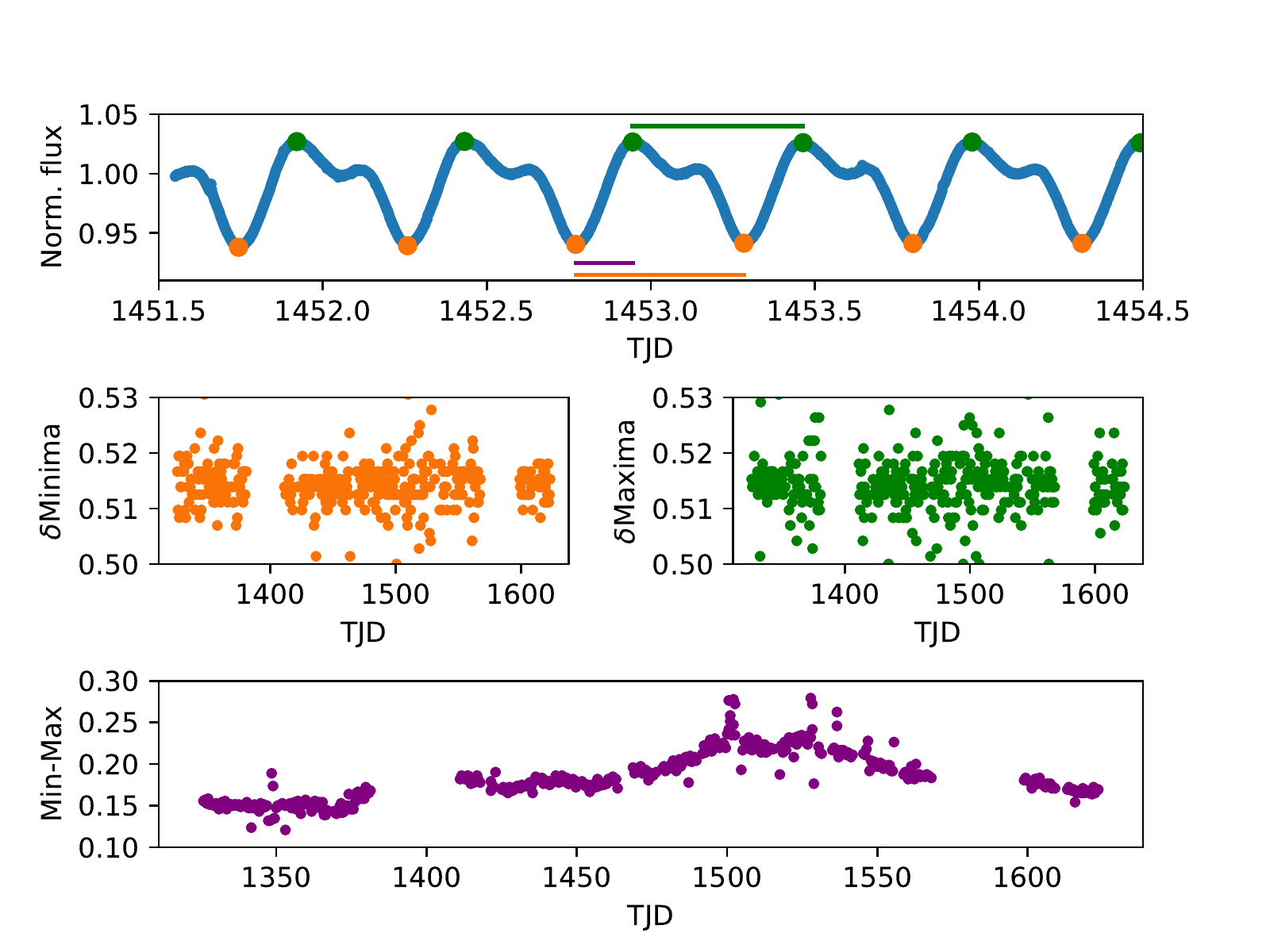}
   \caption{Min-max measurements. One can see a definition of the terms minima (orange), their difference $\delta$min (orange line), maxima (green), their difference $\delta$max (green line), and the difference min-max (purple line); their distributions (middle panels); and the evolution of their difference (bottom panel).       }
     \label{fig:dmaxmin}
  \end{figure}

 \begin{figure}[t] 
  \centering
  \includegraphics[width=\hsize]{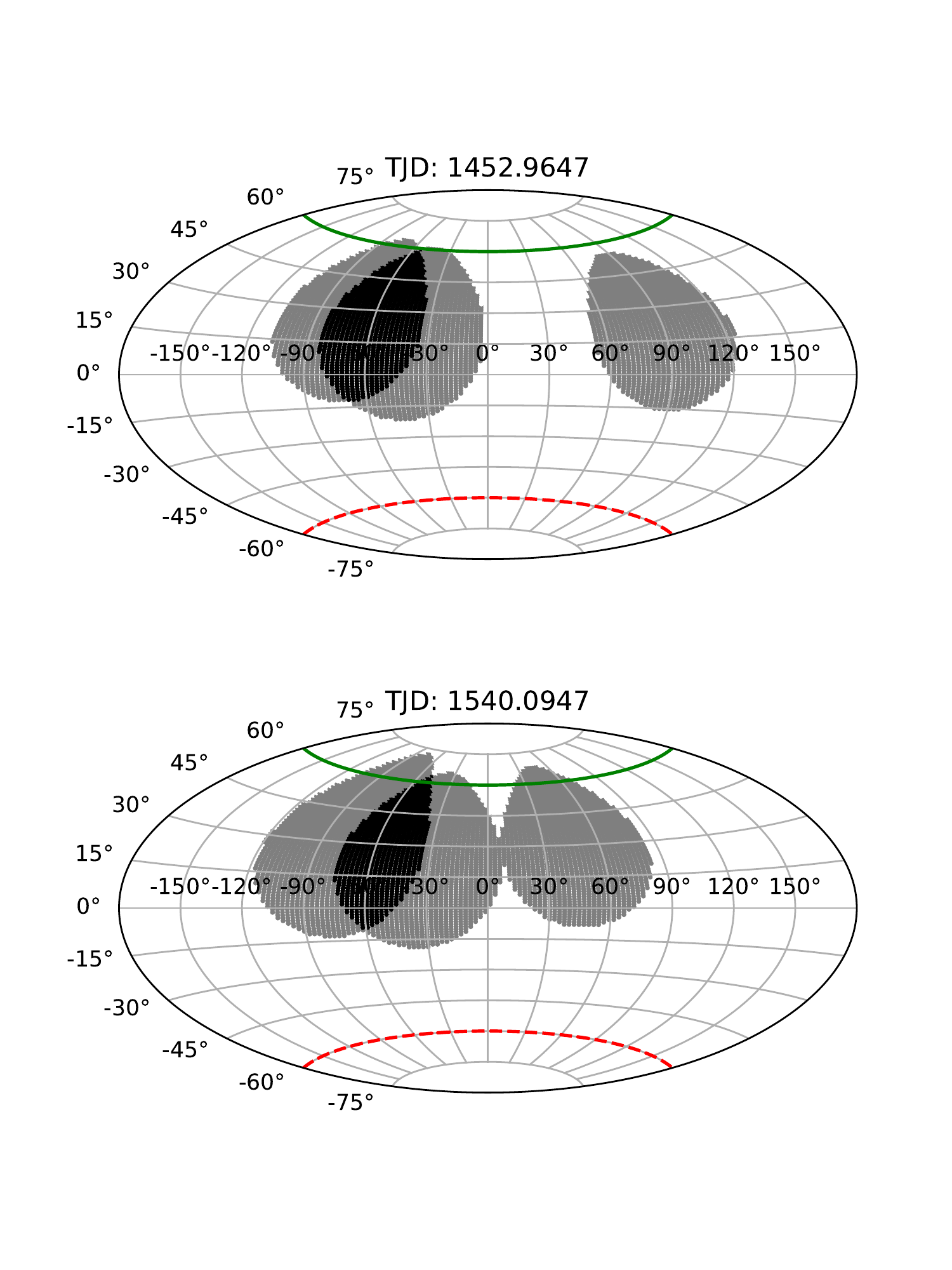}
   \caption{Aitoff projections of the spot configuration on AB Dor at TJD~=~1452.9647 (upper plot) and TJD~=~1540.0947 (lower plot).}
     \label{fig:aitABDor}
  \end{figure}

To connect this result with the phenomenology of the light curve and, given that the contrast between the dark and bright areas in Fig.~\ref{fig:0514275br} (i.e., the peak-to-peak variation of the light curve) reaches its maximum value between TJD $\simeq$ 1500 days and TJD $\simeq$ 1580 days\footnote{TJD is the TESS JD time, which is defined as the difference between the barycentric dynamical time minus the number 2 457 000.}, we also measured the average peak-to-peak variation of the same parts of the light curve, which we 
show in the lower panel of Fig.~\ref{fig:p2pgls}.\ We estimate the typical error for those measurements to be
on the order of 10$^{-4}$

At first sight, the GLS results and the peak-to-peak variation of the light curve, as presented in 
Fig.~\ref{fig:p2pgls}, show no direct correlation between them or between the GLS results and the phenomenology of the light curve (i.e., the relative ``motion'' of the dark areas in Fig.~\ref{fig:0514275br}). Given the possible inadequacy of the GLS periodogram to describe a light curve that can hardly be described as sinusoidal due to its continually evolution, we attempted to measure the period of the modulations using yet another method, that is, to measure the time distance between consecutive minima and maxima, to which we refer to as min-max in the following.\ Again, we estimate the
typical uncertainty of those measurements to be on the order of one minute.
 
In Fig.~\ref{fig:dmaxmin} we show the results of our min-max measurements. The upper panel is explanatory and shows what we mean by the terms minima (orange), maxima (green), and their difference (purple), using a part of the actual TESS light curve of AB Dor.\ We note that the precision of the measurements is so large that the error bars are smaller than the dots in this plot. 
As shown in Fig.~\ref{fig:dmaxmin}, the values of $\delta$min = 0.51396 days and $\delta$max = 0.51488 days are stable in time but not equal (see Table \ref{tab:per}), although the difference of their average values appears to be insignificant. 
We also measured the min-max difference (depicted with purple in Fig.~\ref{fig:dmaxmin}) to test how the time difference between minima and maxima varies during the observations of AB~Dor. The results, shown in the lower panel of Fig.~\ref{fig:dmaxmin}, suggest that the measured differences between the consecutive minima and maxima appear to be time-dependent, quantify the phenomenology of the light curve shown in Fig.\ref{fig:0514275br}, and agree with the peak-to-peak variations shown in Fig.~\ref{fig:p2pgls}.

 \subsection{Spot modeling}

The known inclination of AB~Dor allowed us to use a spot model to derive the 
location of the spots on the stellar surface. This type of spot modeling has been applied to photometric light curves
for a long time and the difficulties encountered are succinctly described, for example, by \cite{harmon2000}.
In principal, one is trying to reconstruct a 2D-surface from 1D data (i.e., a light curve), and such problems
are known as ``ill-posed'' in the mathematical literature.  To overcome these difficulties, one either has to
choose some kind of regularization, that is to say surfaces as have to be as ``smooth'' as possible, for example, or one has to
limit the number of free parameters, that is, the number of spots on the surface and hence the
number of free parameters. It is a difficult problem to decide how many free parameters one needs,
in practice, to describe a given light curve. We estimate the number to be in the range of eight to ten from numerical experiments. At any rate, it must be clear that the inherent mathematical difficulties of 
any light curve inversion does not allow for the construction of unique models. 
  
\begin{figure}
  \centering
  \includegraphics[width=\hsize]{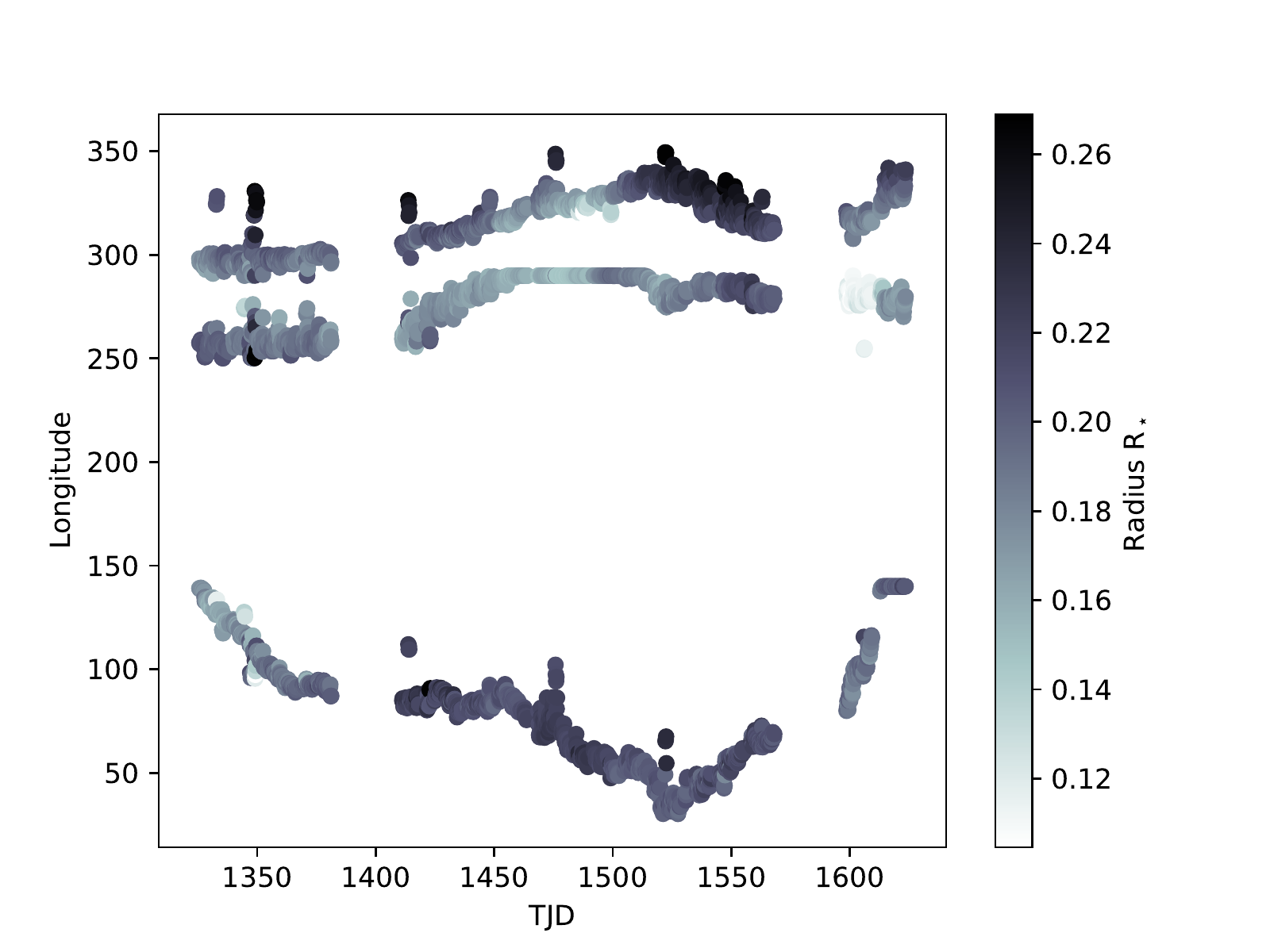}
   \caption{Spot modeling result for the light curve of AB Dor.\ The measured sizes of the spots (in $R_\star$ units) appear in grayscale. 
       }
     \label{fig:long_evo2}
  \end{figure}

To measure the positions of the star spots on AB~Dor, we, therefore, used a 3+2 spot model similar to the one described and used by \cite{Ioannidis2016} with the difference that the spots are not locked onto the stellar equator, but rather they may take latitude values as large as \ang{60}. The number of spots in our spot model
remains constant for modeling each phase of the light curve. Although our model assumes
nonoverlapping circular spots, we think of those areas more like large active regions consisting of many unresolved, 
little spots, rather than a large uniform spot. Furthermore, our spot models assumes a
spot temperature of $t_\mathrm{sp}$ = 0, due to the correlation of $t_\mathrm{sp}$ with the spot radius. 
An impression of a typical modeling result is given in Fig.\ref{fig:aitABDor}, which shows
the modeled spotted surface of AB~Dor at two epochs close 
to the mean and average high peak-to-peak variation in an Aitoff projection.\ The more detailed results of our spot model are shown in Fig.~\ref{fig:long_evo2}, which provides the temporal evolution of
longitudes of the derived spot positions. 
As we show in Fig. \ref{fig:long_evo2}, our model suggests the existence of a spot, separated by approximately \ang{120} by a pair of other spots early in the TESS observations. The angular separation of the spots becomes smaller as time goes by, reaching its smallest value (i.e., $\sim$\ang{100}) between TJD $\simeq$ 1500 days and TJD $\simeq$ 1550 days. Just before the end 
of the TESS observations, the spots appear to move ``back'' to a point close to their original positions.

 \begin{figure}

  \centering
  \includegraphics[width=\hsize]{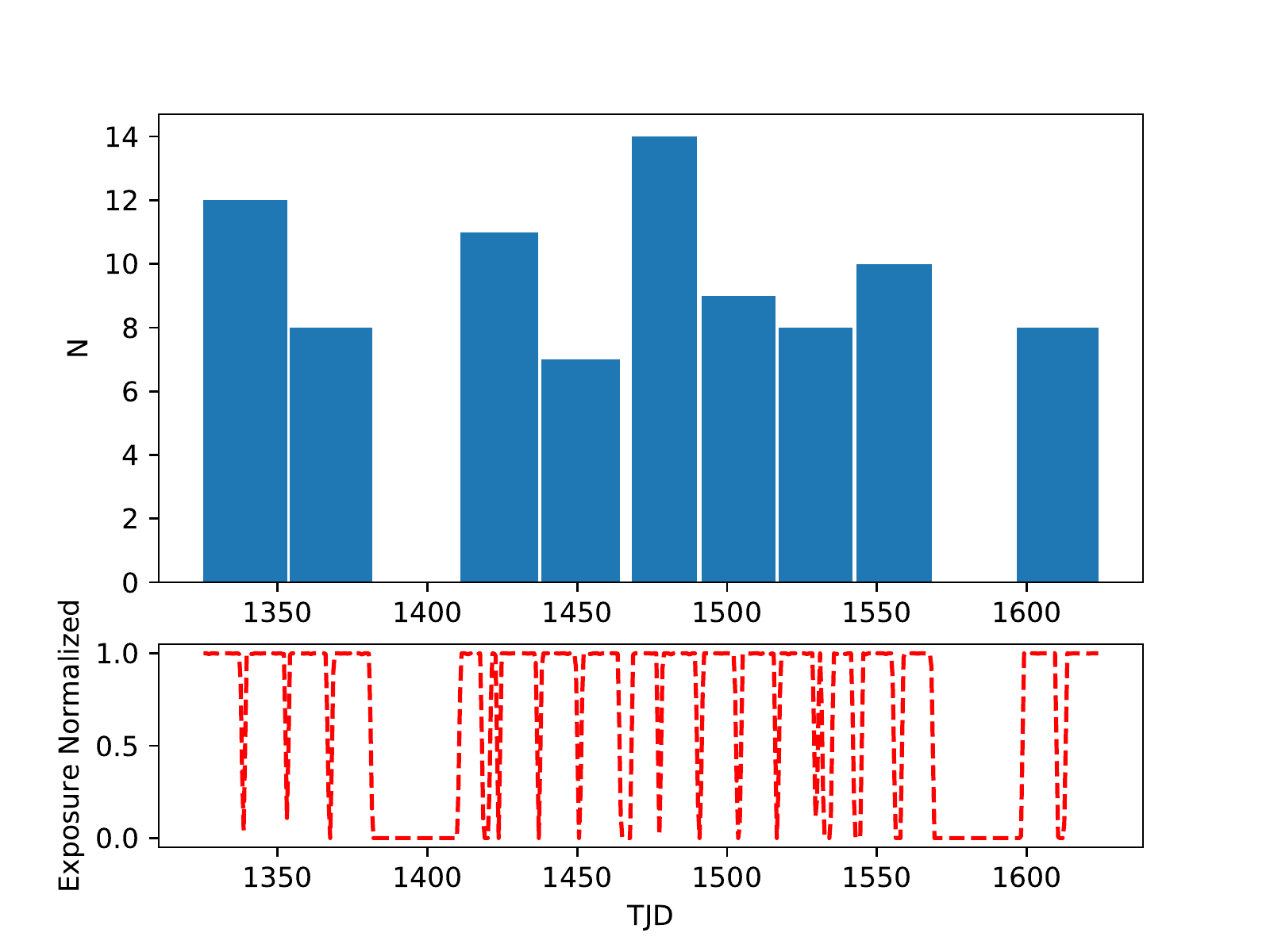}
   \caption{Distribution of ``big'' flares of AB~Dor during the TESS observation time}
     \label{fig:flare_dist}
  \end{figure}

 \begin{figure}
  \centering
  \includegraphics[width=\hsize]{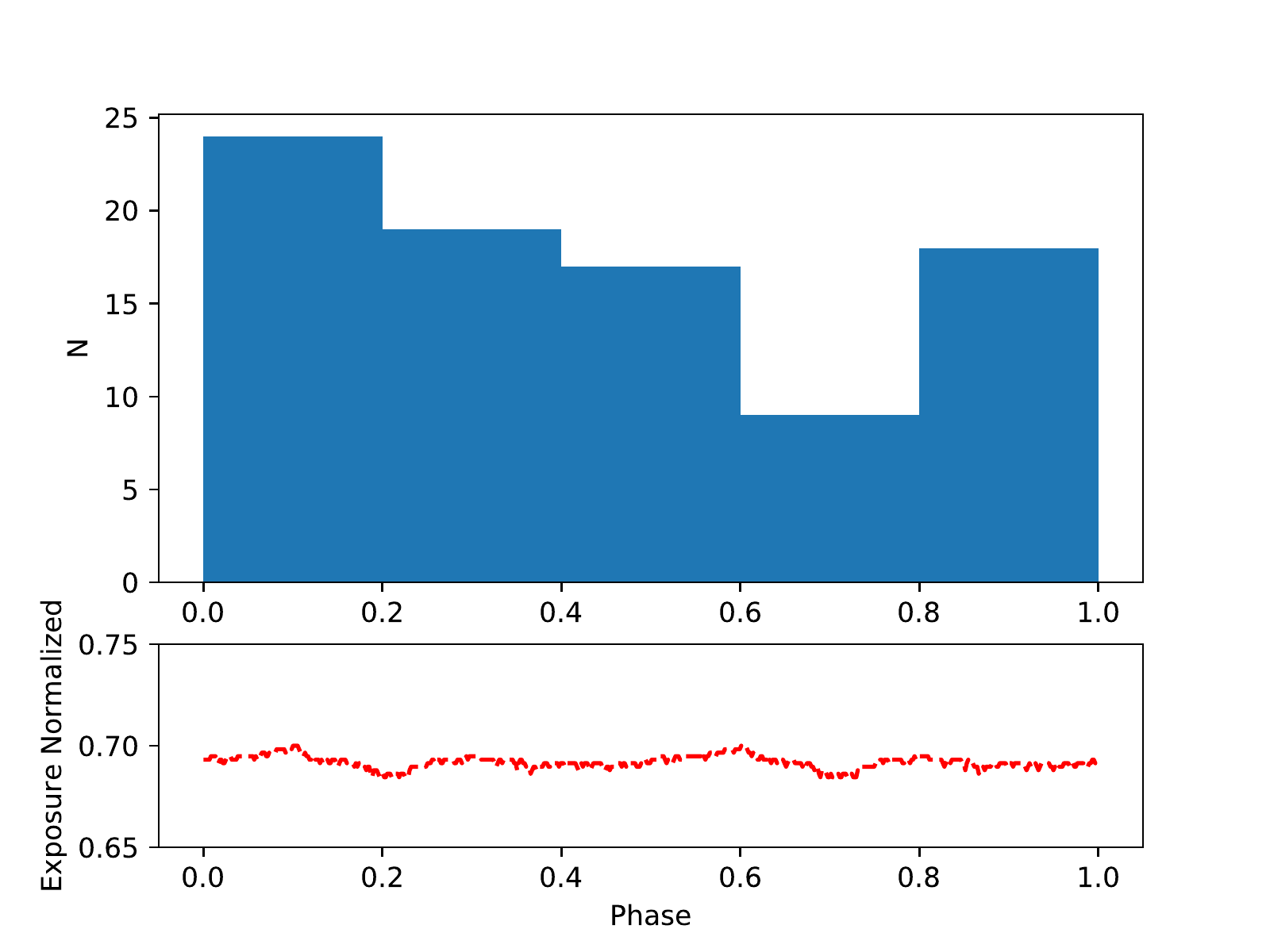}
   \caption{Same as Fig.\ref{fig:flare_dist}, but now phased to the rotation period of $P_\mathrm{rot} = 0.514275$ days.}
     \label{fig:flare_disth}
  \end{figure}

\subsection{Flares}

As is shown by \cite{schmitt2019}, the TESS light curve of AB~Dor shows many ``big'' flares. Although the goal of this study is not to focus specifically on the flares, we used the big flares \cite[as defined by][]{schmitt2019} to measure their temporal distribution. The detection and categorization of the flares was achieved manually. Specifically, we visually inspected the TESS light curve of AB~Dor and identified the flares with peak luminosities larger than $1\times10^{31}$ erg/s (i.e., $\sim$0.5\% in the AB~Dor light curve); as flare occurrence times, we used the time just when the rising phase of the flare began. 

Following this procedure, we manually identified 87 big flares in the TESS light curve of AB~Dor, that is to say 
a little more than two of those flares per week. In Fig.~\ref{fig:flare_dist} and Fig.~\ref{fig:flare_disth}, we show 
the distribution of these big flares as a function of time as well as rotation phase (assuming a rotation 
period of $P_\mathrm{rot} = 0.514275$ days), respectively. The red lines denote the observation coverage rate for both figures, where ``0'' denotes no observations. Although we do not detect significant variations or trends with time,
it becomes clear from Fig.~\ref{fig:flare_disth} that there is a preferential behavior of the
flare occurrence related to the rotation period of the star: There is $\sim$55\% lower major flare occurrence between phase 0.6 and 0.8 with a significantly more than 2$\sigma$ (assuming that the flare occurrence follows a Poisson distribution). In comparison to Fig.~\ref{fig:0514275br}, we note that this is exactly that part of the light curve when the star is less spotted, which is consistent with the view that
there are fewer flares when there are fewer spots. 
 
\section{Discussion and conclusions}

\subsection{Rotation period}
\label{sect:rot}
  
As apparent from Table~\ref{tab:per}, the rotation period of AB Dor is in the range between $P_\mathrm{rot} = 0.51396$ days and $P_\mathrm{rot} = 0.51488$ days.
However, we can obtain a more precise measurement if we assume that the bright parts in the light curves of AB~Dor, that is, the phase range $0.25 < \phi < 0.5$, is not changing its position on the star. Under this assumption, 
we can calculate AB~Dor's rotation period by combining the photometric data from \cite{Kuerster1994} and TESS.
 \cite{Kuerster1994} listed their photometry times, from which we can derive the time of maximum light during their observations as T$_0$ = 2447579.5307. In the TESS light curve, we find one of its maxima at the time T$_1$ = 2458473.0.  
By assuming a certain number of rotations, we can then reproduce the different rotation periods derived in the literature; we show in the third column of Table~\ref{tab:per} these values, which range from 21160 and 21195.
The rotation period derived by \cite{schmitt2019} was also derived from the period between maximum light
epochs, and thus, under our assumption that maximum light occurs at the same rotation phase, some 21181
AB~Dor ``days'' would have elapsed, which leads to a period of $P_\mathrm{rot}$ = 0.51430 days.

\subsection{Spot positions}
\label{sect:positions}

During the whole TESS observation process of AB~Dor, there was not a single moment when the star was spotless, that is, at least in the zone below \ang{60} latitude. AB~Dor does not seem to be special in that manner, and
there are many examples of other stars with similar long-term behavior, yet, remarkably, we seem to have somehow observed the same set of active regions. The similarity of our results, as seen in Fig.~\ref{fig:aitABDor}, with those shown in Figs. 12 to 15 by \cite{Kuerster1994} suggest that the active regions formation on AB~Dor's surface has changed only a little or not at all for the last $\sim$30 years, which is in agreement with our measurements from Sec.~\ref{sect:rot}. 

The results of our spot model analysis suggest that the spots on AB~Dor appear at relatively low latitudes in the range of $\ang{20}<\phi<\ang{30}$. The low latitudes result from the fact that the model needs to account
for the observed substantial peak-to-peak variations. This is not in contradiction with the assumption of polar spots, which would drop the star's total observed flux and only contribute to the rotational modulation on a lower scale.
Although we used zero spot temperatures in our model, the calculated active region sizes are still enormous for solar
standards. Although we cannot accept or reject those results, such spot formation scenarios appear nonphysical, suggesting that there may be alternative explanations for the measured peak-to-peak variation.

AB~Dor is one of the best candidates to apply a spot model, and we can be confident that the longitude results, shown in Fig.~\ref{fig:long_evo2}, reflect the reality. However, if the evolution of the active regions was to follow what Fig.~\ref{fig:long_evo2} suggests, the active regions of AB~Dor should alter their differential rotation patterns (i.e., the rotation period of the active regions would have to change dramatically at TJD$\sim$1525 days). As a result, we think that the ``change in course'' observed at TJD$\sim$1525 is connected with the appearance, evolution, and disappearance of smaller spot groups (parts of the larger active regions) and not a consequence of differential rotation.

During the phases with visible spot features, we see 40\% more flares.\ This implies that (1) the flares occurrence is directly connected to the spot occurrence and (2) 60\% of the flares should occur in areas with constant spot visibility, that is, the polar cap.

\subsection{Spot lifetimes}

Although the formation of the active regions in the photosphere of AB~Dor 
seems to be restricted in longitude, the temporal evolution of the light curve does suggest that we are not 
looking at the same active regions all the time. Rather, the modulations of AB~Dor are created by starspots, which 
are born, evolve, and eventually dissolve around the same general active longitude 
(see Fig.~\ref{fig:0514275br} between phase $\sim$0.25 and $
\sim$0.6). 

To this end, we inspected the GLS periodogram of the TESS AB~Dor light curve, which is also
in the period range above five days and
which identified formally significant peaks in the period range between 11 and 30 days, 
with the most prominent peaks being located around 13 and 26 days.  
However, it is rather difficult to assess 
the true nature of those signals, given the observation strategy of TESS, that is, repointing
the telescope every 27 days, a procedure
which could introduce spurious periodicities in the GLS periodogram.

\begin{figure}
  \centering
  \includegraphics[width=\hsize]{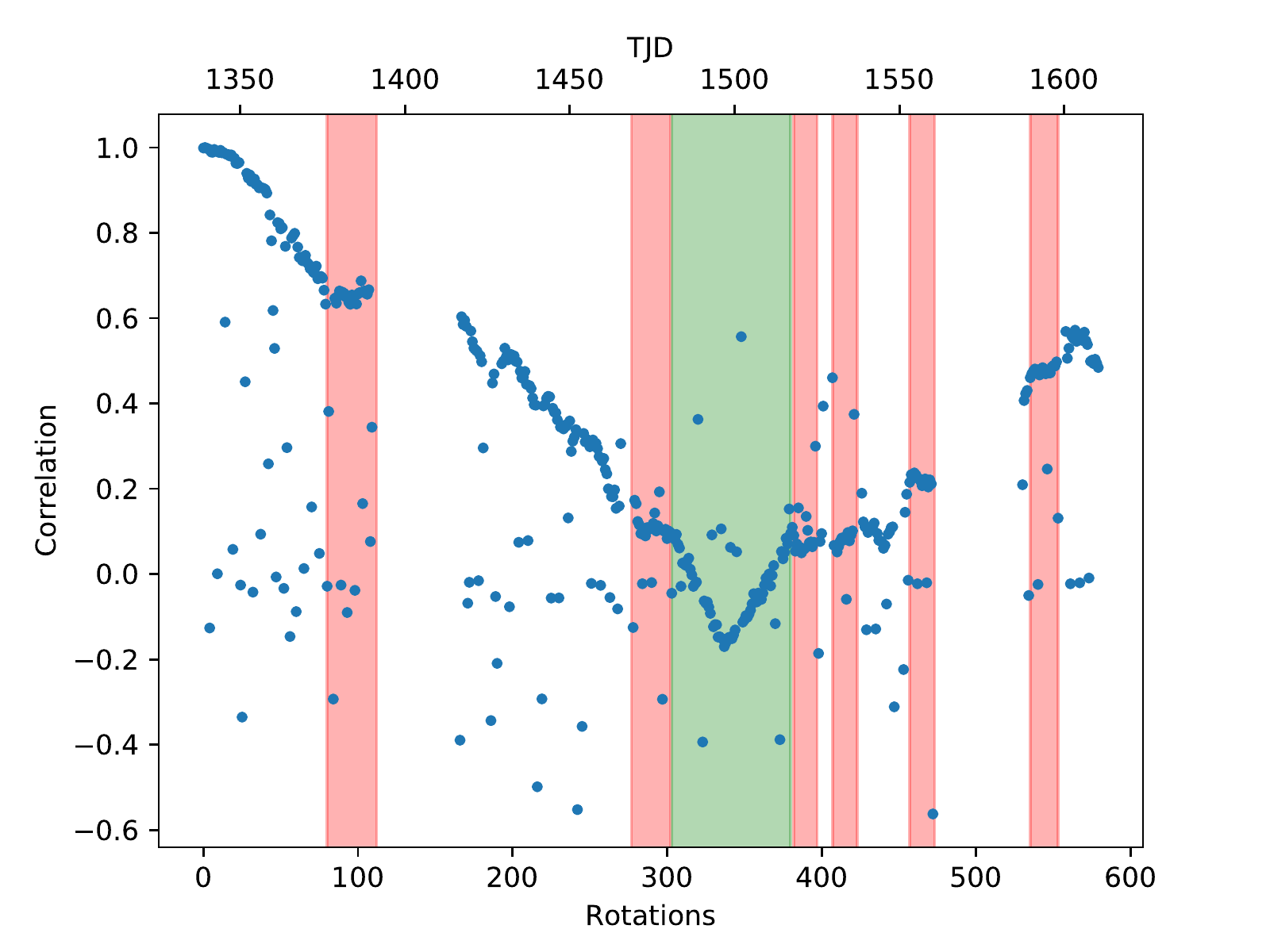}
   \caption{Cross-correlation between the different rotations of AB~Dor TESS light curve, using the first observed rotation in the same data as a reference (see text for details). 
    }
     \label{fig:lc_corr}
  \end{figure}

To better understand the nature of these additional signals in the GLS periodogram of AB~Dor, we 
therefore followed a different approach and investigated the correlation structure of the TESS
AB~Dor light curve. Assuming a rotation period of $P_\mathrm{rot}$ = 0.51488 days, we could
phase the observed light curve of AB~Dor and consider the light curve as a function of rotation or
time. We then specifically cross-correlated the phased light curve of TESS' first rotation of AB~Dor 
with each phased light curve derived from the subsequently observed TESS rotations and
we show the results of this procedure in Fig.~\ref{fig:lc_corr}, that is,
we plotted the computed cross-correlation coefficient of the n$^{th}$ rotation (denoted by time in TJD) with respect to the first rotation.
The points scattering around a correlation coefficient of zero are mainly caused by flares and other systematic errors in the 
light curve, yet the bulk of the data points (i.e., the correlation
coefficients of the phased light curves) shown in Fig.~\ref{fig:lc_corr} follow a clear
pattern of increasing decorrelation until TJD 1500 (i.e., after $\sim$350 rotations), 
from which point the cross-correlation starts increasing again.

  \begin{figure}
  \centering
  \includegraphics[width=\hsize]{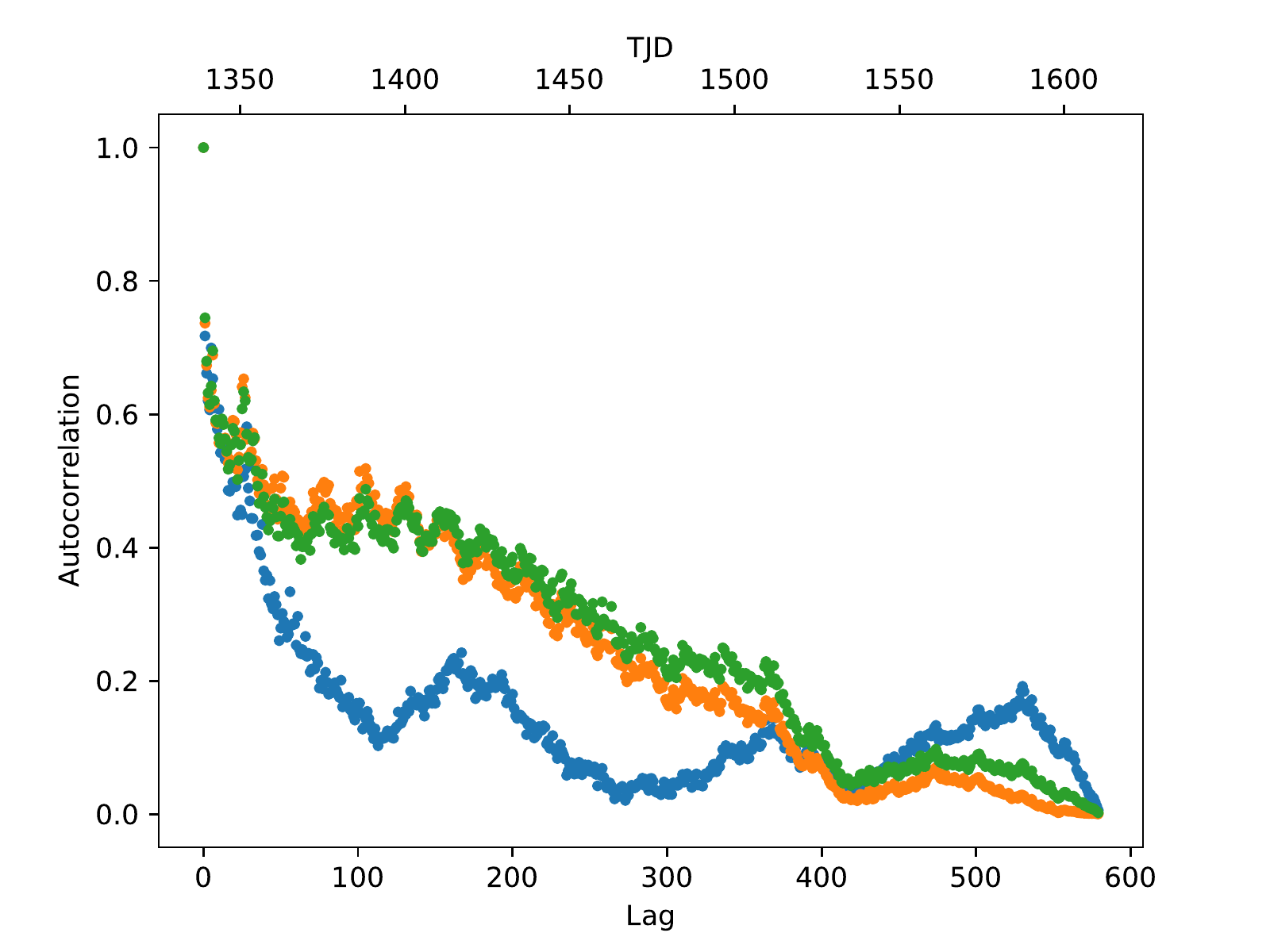}
   \caption{Rows 0, 300 and 600 of the autocorrelation functions A$_{k,j}$ from the AB~Dor light-curve; see text for details.
    }
     \label{fig:test}
  \end{figure}

\begin{figure}
  \centering
  \includegraphics[width=\hsize]{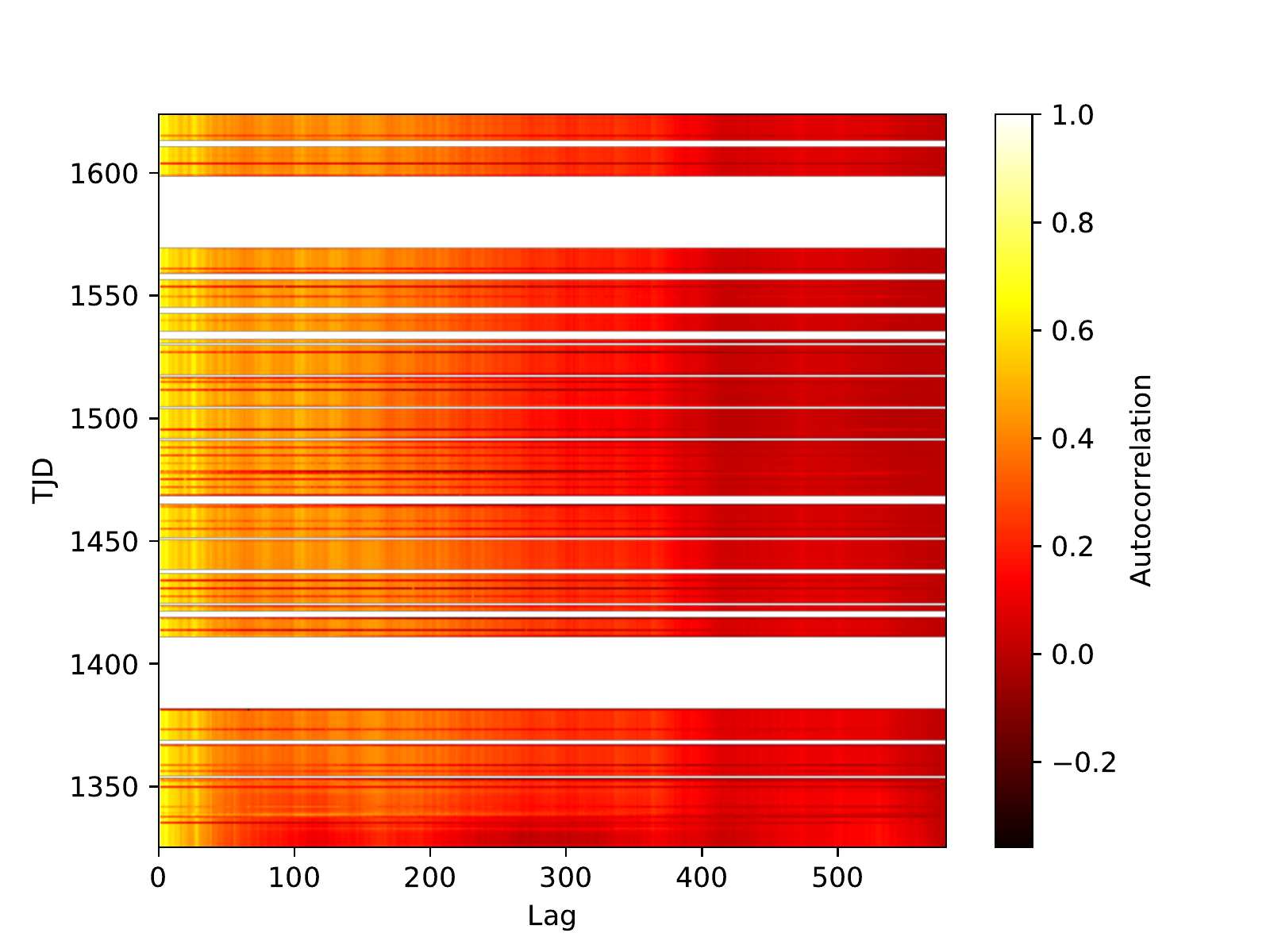}
   \caption{Time-lag diagram of autocorrelation functions A$_{k,j}$.\ It is important to note the decrease in correlation (see text for details).
    }
     \label{fig:lc_acor}
  \end{figure}

\begin{figure}
  \centering
  \includegraphics[width=\hsize]{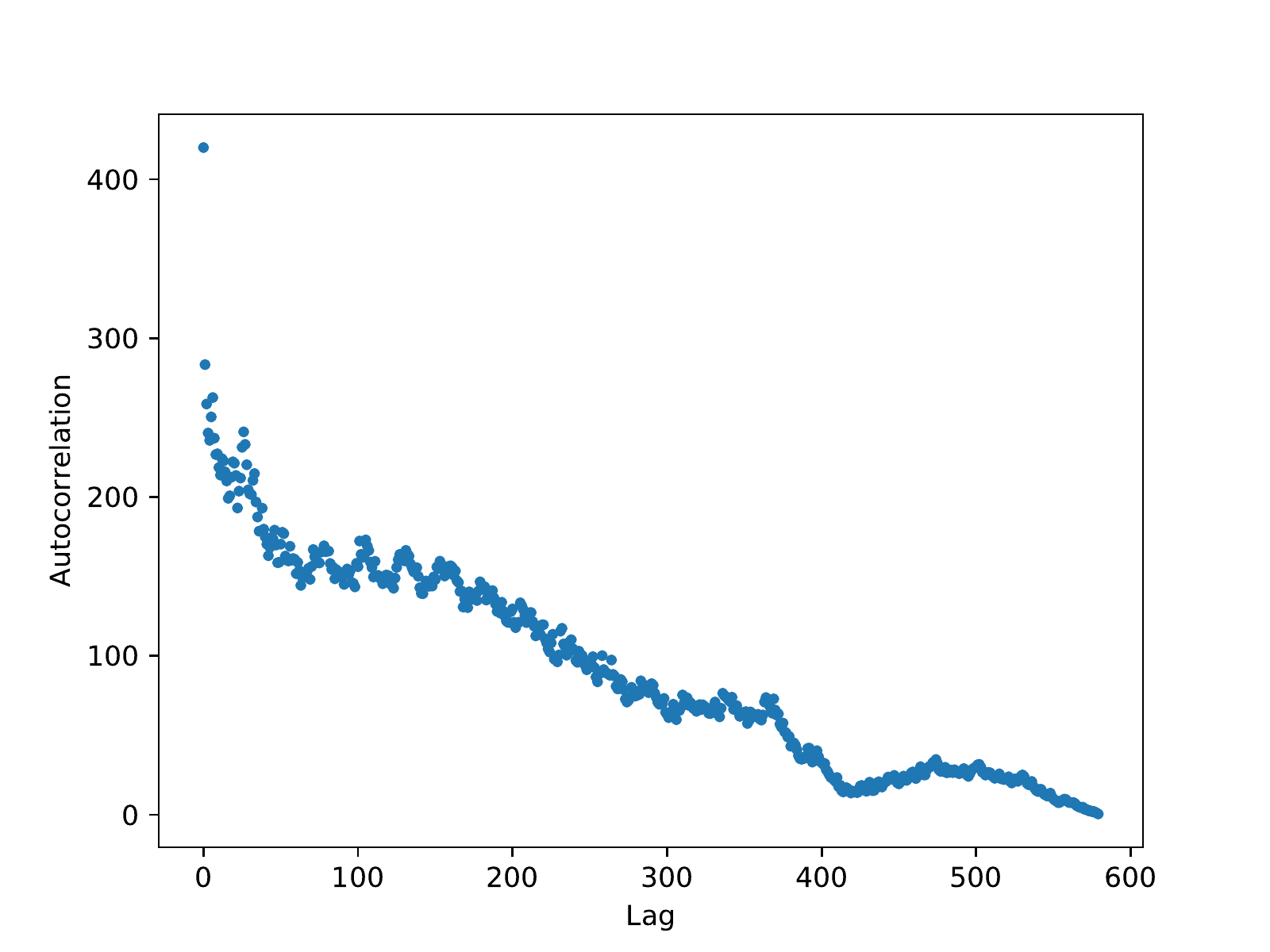}
   \caption{Mean autocorrelation as function of lag for the light curve of AB~Dor; see text for details.
    }
     \label{fig:lc_acor_sum}
  \end{figure}

The choice of the first observed rotation as a reference is, of course, arbitrary, and in fact, we can choose any rotation 
as a reference. More generally, we can compute a quadratic matrix C$_{i,j}$, which contains, as its elements, the 
cross-correlation coefficients computed between any of the observed rotations $i = t_{0},...,t_{N}$, where $t_{i}$ 
denotes the central time of the $i^{th}$ rotation, and the light curve of the $j^{th}$ rotation as a reference; thus, 
by construction, C$_{i,i}$ = 1, for i =1,N, where N denotes the total number of rotations.\ To provide an example, the 
correlation coefficients shown in Fig.~\ref{fig:lc_corr} refer to C$_{1,j}$.
Next, we investigated the process of decorrelation by determining the autocorrelation
of the computed cross-correlation functions. Thus for the cross-correlation C$_{i,j}$ (with the index $i$ denoting 
time and the index $j$ denoting the chosen reference rotation), we computed correlation coefficients A$_{k,j}$, with the 
index $k$ denoting the chosen lag (in terms of rotations) and the index $j$ again denoting the chosen 
reference rotation; hence by construction we have A$_{0,j}$ = 1.
In Fig.~\ref{fig:test} we show a few examples of such autocorrelation functions, namely those using rotation 1 as a reference
(blue), rotation 300 (orange), and rotation 580 (green). In all of these functions, a ``wiggly'' structure with a period of about
26 rotations (i.e., around $\sim$ 13 days) appears. In Fig.~\ref{fig:lc_acor}, we show -- as a time-lag diagram -- the full array of
autocorrelations, with the x-axis displaying the lag and the y-axis showing the chosen reference rotation. In all functions, A$_{k,j}$, which is a 
sequence of vertical patterns, appears, which reflects the ``wiggly'' structure in Fig.~\ref{fig:test}. To obtain results 
independent of the reference rotation, we ``collapsed'' the matrix A$_{k,j}$ along the y-axis, that is, 
computing the mean autocorrelation function $S(k) = {1\over n} \sum_{j=1}^{N} A_{k,j}$ to finally arrive at
Fig.~\ref{fig:lc_acor_sum}, which again shows periodic modulations. It is, of course, suggestive of interpreting these modulations as signatures of spot lifetimes, and we note that at a period of 13~days, a peak in the GLS periodogram is found.

\subsection{Spot simulations}

To better understand the statistical properties of our TESS AB~Dor light curve,
we also simulated the photometric light curve of a star with the physical characteristics of 
AB~Dor, that is, with active longitudes as they appear in Fig.~\ref{fig:0514275br}, 
with constant spot coverage (see Sect.~\ref{sect:positions}), and with
similar temporal evolution as shown in Fig.~\ref{fig:lc_corr}. We note that
the goal of this exercise is not to recreate the exact spot configuration on AB~Dor, but rather 
to investigate whether (a) there is any spot configuration which would produce the pattern apparent in 
Fig.~\ref{fig:lc_acor_sum} without assuming an unrealistic light curve and (b) how  
the same pattern is related to the spot lifetimes, while at the same time
keeping our model as simple as possible. 

Via trial and error, we found that the simplest spot configuration to mimic a light curve satisfying our criteria 
requires a combination of five spots, that is, two spots at rather high latitudes (with $R_\mathrm{sp}~/~R_\star=0.05$ 
and \ang{70} latitudes) -- with their larger parts in the continuously visible area of the star -- and a 
set of three spots at lower latitudes (with $0.05 \leq R_\mathrm{sp}~/~R_\star \leq 0.1$), grouped in an area with similar longitudes and latitude values around \ang{20}. In this fashion, we obtain a light curve without any flat parts, that is, at least one spot
is always present on the visible hemisphere -- a configuration which is in agreement with previous Doppler imaging studies on AB~Dor (see, \citet{donati_1997} and \citet{jeffers_2007}). Next, we assumed solar-like-differential rotation following
\cite{donati_1997} to create the needed temporal evolution of the light curve and set the rotation periods for the high and low latitude spots equal to the maximum and minimum values listed in Table~\ref{tab:per}, respectively.  
To create lifetime effects, we assumed that the low latitude spots have similar but not exactly equal lifetimes 
(the mean value of which we set to 6, 12, and 24 days) in different simulations, while the high latitude spots remain constant. 
The difference in the lifetimes of the spots in a given spot group was set
to 0.3 days (e.g., 11.7, 12, and 12.3 days). Since the appearance and disappearance of
spots -- based on our knowledge from solar spots -- is not instantaneous, we describe these effects with an ad hoc ``smoothing function'' and assume that the appearance and disappearance are described by half of a 
cosine wave via the function $R'_\mathrm{spot}$ = $R_\mathrm{spot}$ $\times \, (cos(\mathrm{T}+\pi)+1)/2$ during growth and $R'_\mathrm{spot}$ = $R_\mathrm{spot}$ $\times \, (cos(\mathrm{T})+1)/2$ during the decay, where the dimensionless time $T$ is the ratio between actual time $t$ and the spot lifetime t$_\mathrm{sp}$
via T = t~(2$\pi \, \mathrm{t}_\mathrm{sp}$)$^{-1}$. In our simulations,
the growth or the decay of the spots was arbitrarily set to last 2.5 days.
After the disappearance of a spot, a new spot with the same characteristics replaced it immediately in the same position. 
To avoid sampling biases, we
kept the total number of days in our simulated light curve equal to the observation time.
Additionally, for simplicity, the maximum size of all spots remained the same during the total
time of the simulated light curve, and it was set so the peak-to-peak of the variations  
matched the real AB~Dor light curve.
In this fashion, we generated synthetic light curves choosing spot lifetimes of 6, 12, and 24 days,
which were then subject to the same analysis as the real AB Dor TESS light curves.

\begin{figure}
  \centering
  \includegraphics[width=\hsize]{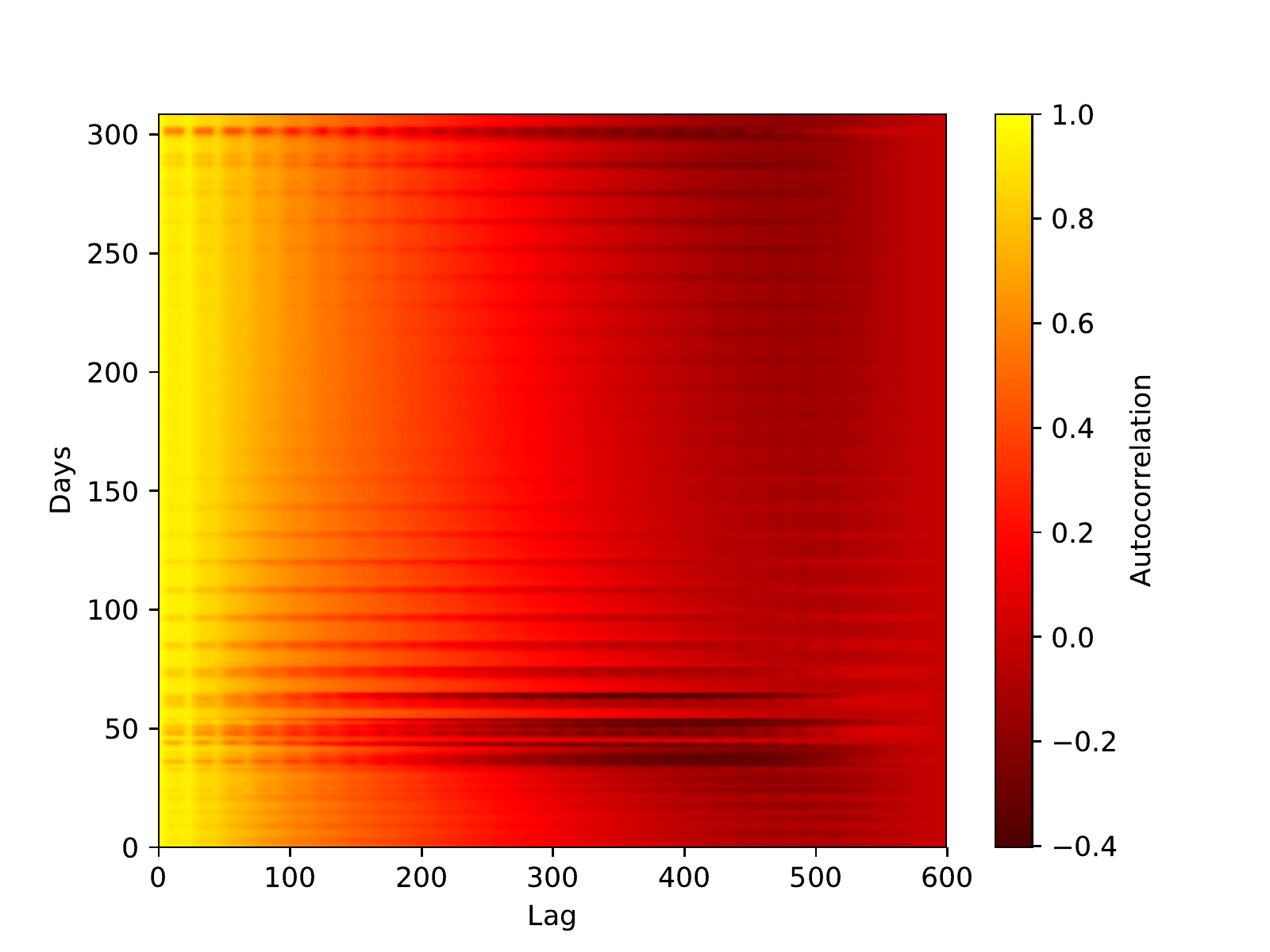}
   \caption{Same as Fig.~\ref{fig:lc_acor}, but from our simulated light curve.
    }
     \label{fig:sim_acor}
  \end{figure}

\begin{figure}
   \centering
  \includegraphics[width=\hsize]{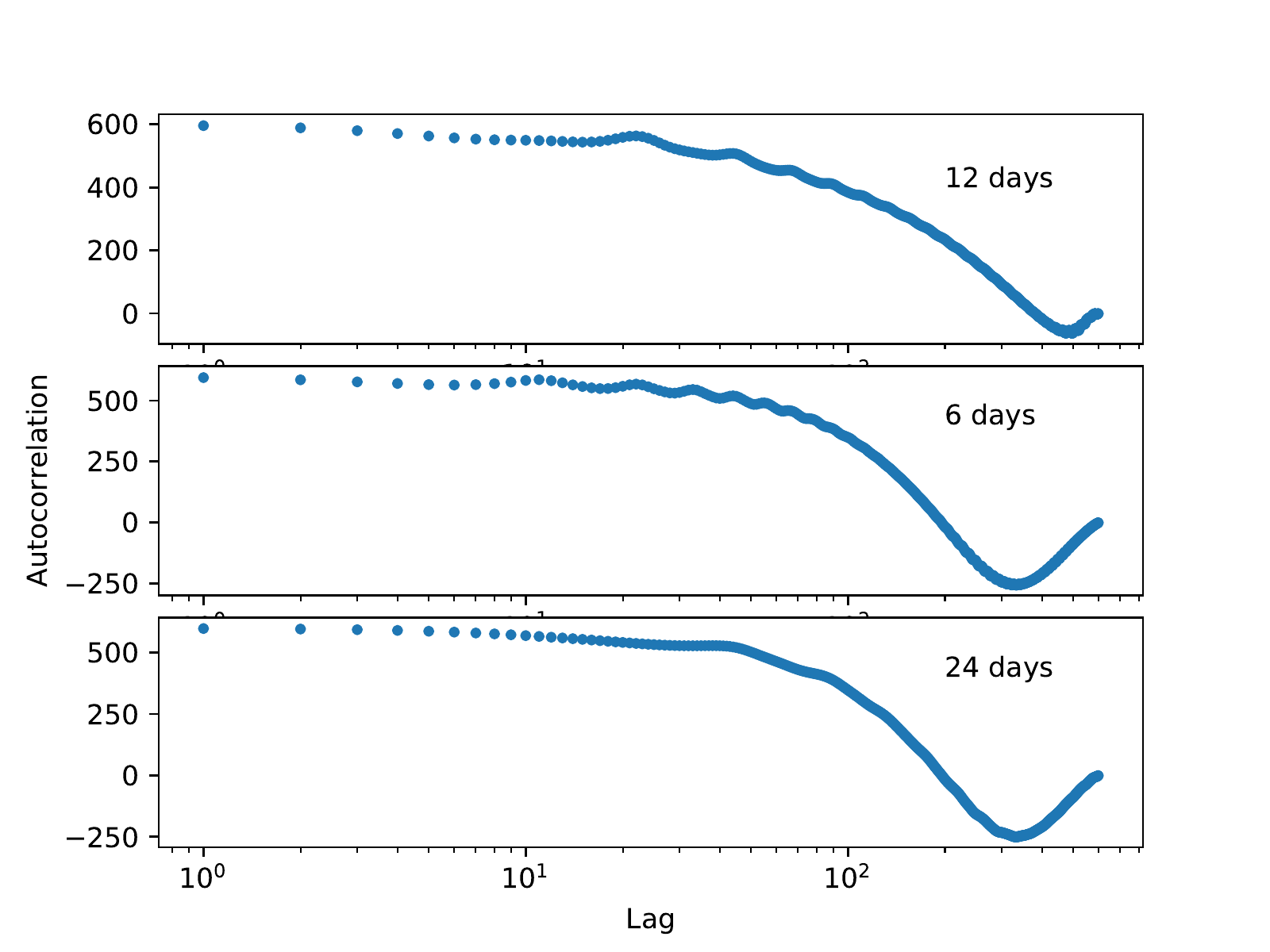}
   \caption{Top panel: Same as Fig.~\ref{fig:lc_acor_sum}, but from our simulated light curve and assumed spot lifetime of 12 days. Middle and bottom panel: Collapsed A$_{ij}$ from our simulated light curves for spots with assumed lifetimes of 6 and 24 days, respectively.
    }
     \label{fig:sim_acor_61224}
  \end{figure}

The time-lag diagram of our simulated light curves analogous to Fig.~\ref{fig:lc_acor} is 
shown in Fig.~\ref{fig:sim_acor}. Although there is great similarity between our simulations and the actual data from AB~Dor, our case, that is to say that the peaks in the A$_{i,j}$ are the result of spot lifetimes,
becomes more clear when we compare the A$_{k,j}$ from simulated light curves with various spot lifetimes. 
In Fig.~\ref{fig:sim_acor_61224} we show the collapsed A$_{k,j}$ values for spots with lifetimes of 
6, 12, and 24 days. 

\subsection{Bright spots}

As discussed in sect.~\ref{sect:positions}, the peak-to-peak variations in the AB~Dor light curve suggest the size of the active regions to be very large, that is, the average spot size from our modeling is R$_\mathrm{spot}$~/~R$_\mathrm{\star}$$\simeq$ 0.18. On top of that, the temperature contrast between those regions and 
the stellar photosphere should be almost maximum (i.e., for our models, we assume T$_\mathrm{spot}$~/~T$_\mathrm{\star}
$ = 0). Although we are not in a position to confirm or disprove of this
hypothesis, in this section, we explore an alternative hypothesis, that is, the existence of bright regions in combination
with ``traditional'' cool active regions. As we know from the Sun, during those periods of the 
11-year cycle when it is more active and has more spots, it is also brighter in total solar irradiance
due to the contribution of plage regions.
Following a similar line of thought, we assume ad hoc that a similar process occurs in 
AB~Dor, and, by using our spot model algorithm, we attempted to reproduce the observed modulations in the TESS light curves of AB~Dor with both bright and dark spots as well as with bright spots alone. 
Although it is possible to replicate the modulations following both assumptions, the problem of the large active regions with high contrast remains. Specifically, it was not possible to 
recreate the activity modulations unless we used large (R$_\mathrm{spot}$~/~R$_\mathrm{\star}$$\simeq$ 0.1), cool 
(T$_\mathrm{spot}$~/~T$_\mathrm{\star}$$\simeq$ 0.3), and bright (T$_\mathrm{spot}$~/~T$_\mathrm{\star}$$\simeq$ 1.7).
As a result, we conclude that assuming the presence of bright, active regions does not lead to
less extreme values for the spot sizes and temperatures.

\subsection{Differential rotation}

As discussed in Sect.~\ref{sect:abdor}, ever since the ``discovery'' of AB~Dor, its
differential rotation of AB~Dor has been the subject of quite a few studies. From their Doppler images, \cite{donati_1997} derived a rotation ``law'' for AB~Dor 
of the form $\Omega(l) = 12.2434 - 0.0564 sin^2(l) $ (rad/day), where $l$ denotes 
the stellar latitude, implying that the polar regions rotate more slowly than equatorial ones.
Evaluating $\Omega(l)$ at a latitude of 30$^{\circ}$ yields a period of 0.513781~days, which is shorter than any of the periods derived from photometry (cf., Tab.~\ref{tab:per}).  
Similarly, analyzing Doppler images of AB~Dor between 1988 and 1994, \cite{jeffers_2007} reported
spectroscopic periods of 0.5142~d, 0.5139~d, and 0.5141~d for features at ``effective''
latitudes of \ang{55.61}, \ang{54.45}, and \ang{56.78}, which translate into equatorial rotation periods
of 0.5134~d, 0.5129~d, and 0.5132~d, respectively. Using photometry taken over 20~years 
and the Occamian spot inversion techniques, \cite{jaervinen_2005} estimated spot periods in the range of 
0.51470~ and 0.51487~d, which from their spot inversion models translates into simultaneously existing spot groups in high (>\ang{50}) and low (<\ang{30}) latitudes. Thus, in comparison, our photometric TESS results agree with the results of \cite{jaervinen_2005}.

\section{Summary}

In this paper, we present the results of one year of almost continuous high precision photometry of 
AB~Dor with TESS, with a focus on the photospheric activity of the star. We measured the rotation period of the star using a variety of methods. Although our measurements of the rotation period of 
AB~Dor differ between themselves as well as with past measurements (see Table~\ref{tab:per}), the difference
is minimal ($\lesssim$~1 min). We find that the rotation period which fits the data best is 
$P_\mathrm{rot} = 0.514275$ days \citep{schmitt2019}. 

The rotation period of the star is stable during the TESS observation time. However, there appears to be
some ``movement'' of spots, which manifests itself as differences in the distances between the minima and maxima of the activity modulations and is also visible in our spot model results. We interpret this movement to be due to spot evolution and differential rotation; the spots are not formed uniformly over the stellar surface and live only in highly preferred longitudinal regions.
There is even evidence that the locations and the configuration of these active regions is similar to
the ground-based observations presented by \cite{Kuerster1994} more than 40 years ago. This of course does not mean that the spots on AB~Dor live
that long. In contrast, we argue through a detailed study of the light curve's correlation structure that typical
spot lifetimes range from between 10 and 20 days. In this data set, we do not see evidence of flip flop as reported by \cite{jaervinen_2005}.

Furthermore, AB~Dor produces frequent flares; we note that we are likely to see only half the produced flares.  When we see the hemisphere of the stars with low spottedness, there is
a $\sim$60\% lower occurrence rate of big flares as compared to the case when the other hemisphere of
the star is visible, confirming the straightforward picture that at least the big flares tend to occur in more heavily spotted areas. 
The fact that the flare occurrence rate does not drop to zero during the times that we ``see'' the unspotted part of the star may suggest that those flares are being produced in the ever-seen active regions close to the poles.  This behavior is, of course, un-solar-like.  While it usually is not possible to locate the flaring region on an unresolved stellar surface, the giant flare on Algol observed and described by \cite{schmitt1999}
did occur in the polar region and might present a precedent for AB~Dor.

\begin{acknowledgements}

This paper includes data collected by the TESS mission, which are publicly available from the Mikulski Archive for Space Telescopes (MAST).

PI acknowledges funding through the DFG grant SCHM 1032/65-1.

Furthermore, we would like to thank the anonymous referee for their comments, which helped improve our paper.

\end{acknowledgements}

\bibliographystyle{aa}
\bibliography{references}

\begin{thebibliography}{22}
\expandafter\ifx\csname natexlab\endcsname\relax\def\natexlab#1{#1}\fi

\bibitem[{{Auvergne} {et~al.}(2009){Auvergne}, {Bodin}, {Boisnard}, {Buey},
  {Chaintreuil}, {Epstein}, {Jouret}, {Lam-Trong}, {Levacher}, {Magnan},
  {Perez}, {Plasson}, {Plesseria}, {Peter}, {Steller}, {Tiph{\`e}ne}, {Baglin},
  {Agogu{\'e}}, {Appourchaux}, {Barbet}, {Beaufort}, {Bellenger}, {Berlin},
  {Bernardi}, {Blouin}, {Boumier}, {Bonneau}, {Briet}, {Butler}, {Cautain},
  {Chiavassa}, {Costes}, {Cuvilho}, {Cunha-Parro}, {de Oliveira Fialho},
  {Decaudin}, {Defise}, {Djalal}, {Docclo}, {Drummond}, {Dupuis}, {Exil},
  {Faur{\'e}}, {Gaboriaud}, {Gamet}, {Gavalda}, {Grolleau}, {Gueguen},
  {Guivarc'h}, {Guterman}, {Hasiba}, {Huntzinger}, {Hustaix}, {Imbert},
  {Jeanville}, {Johlander}, {Jorda}, {Journoud}, {Karioty}, {Kerjean},
  {Lafond}, {Lapeyrere}, {Landiech}, {Larqu{\'e}}, {Laudet}, {Le Merrer},
  {Leporati}, {Leruyet}, {Levieuge}, {Llebaria}, {Martin}, {Mazy}, {Mesnager},
  {Michel}, {Moalic}, {Monjoin}, {Naudet}, {Neukirchner}, {Nguyen-Kim},
  {Ollivier}, {Orcesi}, {Ottacher}, {Oulali}, {Parisot}, {Perruchot},
  {Piacentino}, {Pinheiro da Silva}, {Platzer}, {Pontet}, {Pradines},
  {Quentin}, {Rohbeck}, {Rolland}, {Rollenhagen}, {Romagnan}, {Russ}, {Samadi},
  {Schmidt}, {Schwartz}, {Sebbag}, {Smit}, {Sunter}, {Tello}, {Toulouse},
  {Ulmer}, {Vandermarcq}, {Vergnault}, {Wallner}, {Waultier}, \&
  {Zanatta}}]{auvergne_2009}
{Auvergne}, M., {Bodin}, P., {Boisnard}, L., {et~al.} 2009, \aap, 506, 411

\bibitem[{{Borucki} {et~al.}(2010){Borucki}, {Koch}, {Basri}, {Batalha},
  {Brown}, {Caldwell}, {Caldwell}, {Christensen-Dalsgaard}, {Cochran},
  {DeVore}, {Dunham}, {Dupree}, {Gautier}, {Geary}, {Gilliland}, {Gould},
  {Howell}, {Jenkins}, {Kondo}, {Latham}, {Marcy}, {Meibom}, {Kjeldsen},
  {Lissauer}, {Monet}, {Morrison}, {Sasselov}, {Tarter}, {Boss}, {Brownlee},
  {Owen}, {Buzasi}, {Charbonneau}, {Doyle}, {Fortney}, {Ford}, {Holman},
  {Seager}, {Steffen}, {Welsh}, {Rowe}, {Anderson}, {Buchhave}, {Ciardi},
  {Walkowicz}, {Sherry}, {Horch}, {Isaacson}, {Everett}, {Fischer}, {Torres},
  {Johnson}, {Endl}, {MacQueen}, {Bryson}, {Dotson}, {Haas}, {Kolodziejczak},
  {Van Cleve}, {Chandrasekaran}, {Twicken}, {Quintana}, {Clarke}, {Allen},
  {Li}, {Wu}, {Tenenbaum}, {Verner}, {Bruhweiler}, {Barnes}, \&
  {Prsa}}]{borucki_2010}
{Borucki}, W.~J., {Koch}, D., {Basri}, G., {et~al.} 2010, Science, 327, 977

\bibitem[{{Collier Cameron} \& {Donati}(2002)}]{cameron_2002}
{Collier Cameron}, A. \& {Donati}, J.~F. 2002, \mnras, 329, L23

\bibitem[{{Donati} \& {Collier Cameron}(1997)}]{donati_1997}
{Donati}, J.~F. \& {Collier Cameron}, A. 1997, \mnras, 291, 1

\bibitem[{{Donati} {et~al.}(2003){Donati}, {Collier Cameron}, {Semel},
  {Hussain}, {Petit}, {Carter}, {Marsden}, {Mengel}, {L{\'o}pez Ariste},
  {Jeffers}, \& {Rees}}]{donati2003}
{Donati}, J.~F., {Collier Cameron}, A., {Semel}, M., {et~al.} 2003, \mnras,
  345, 1145

\bibitem[{{Fr{\"o}hlich}(2012)}]{froehlich_2012}
{Fr{\"o}hlich}, C. 2012, Surveys in Geophysics, 33, 453

\bibitem[{{Guirado} {et~al.}(2011){Guirado}, {Marcaide}, {Mart{\'\i}-Vidal},
  {Le Bouquin}, {Close}, {Cotton}, \& {Montalb{\'a}n}}]{guirado_2011}
{Guirado}, J.~C., {Marcaide}, J.~M., {Mart{\'\i}-Vidal}, I., {et~al.} 2011,
  \aap, 533, A106

\bibitem[{{Harmon} \& {Crews}(2000)}]{harmon2000}
{Harmon}, R.~O. \& {Crews}, L.~J. 2000, \aj, 120, 3274

\bibitem[{{Innis} {et~al.}(1988{\natexlab{a}}){Innis}, {Coates}, \&
  {Thompson}}]{innis_1988a}
{Innis}, J.~L., {Coates}, D.~W., \& {Thompson}, K. 1988{\natexlab{a}}, \mnras,
  233, 887

\bibitem[{{Innis} {et~al.}(1988{\natexlab{b}}){Innis}, {Thompson}, {Coates}, \&
  {Evans}}]{innis_1988b}
{Innis}, J.~L., {Thompson}, K., {Coates}, D.~W., \& {Evans}, T.~L.
  1988{\natexlab{b}}, \mnras, 235, 1411

\bibitem[{{Ioannidis} \& {Schmitt}(2016)}]{Ioannidis2016}
{Ioannidis}, P. \& {Schmitt}, J.~H.~M.~M. 2016, \aap, 594, A41

\bibitem[{{J{\"a}rvinen} {et~al.}(2005){J{\"a}rvinen}, {Berdyugina},
  {Tuominen}, {Cutispoto}, \& {Bos}}]{jaervinen_2005}
{J{\"a}rvinen}, S.~P., {Berdyugina}, S.~V., {Tuominen}, I., {Cutispoto}, G., \&
  {Bos}, M. 2005, \aap, 432, 657

\bibitem[{{Jeffers} {et~al.}(2007){Jeffers}, {Donati}, \& {Collier
  Cameron}}]{jeffers_2007}
{Jeffers}, S.~V., {Donati}, J.~F., \& {Collier Cameron}, A. 2007, \mnras, 375,
  567

\bibitem[{{Jenkins} {et~al.}(2016){Jenkins}, {Twicken}, {McCauliff},
  {Campbell}, {Sanderfer}, {Lung}, {Mansouri-Samani}, {Girouard}, {Tenenbaum},
  {Klaus}, {Smith}, {Caldwell}, {Chacon}, {Henze}, {Heiges}, {Latham},
  {Morgan}, {Swade}, {Rinehart}, \& {Vanderspek}}]{Jenkins2016}
{Jenkins}, J.~M., {Twicken}, J.~D., {McCauliff}, S., {et~al.} 2016, in Society
  of Photo-Optical Instrumentation Engineers (SPIE) Conference Series, Vol.
  9913, \procspie, 99133E

\bibitem[{{Kuerster} {et~al.}(1994){Kuerster}, {Schmitt}, \&
  {Cutispoto}}]{Kuerster1994}
{Kuerster}, M., {Schmitt}, J.~H.~M.~M., \& {Cutispoto}, G. 1994, \aap, 289, 899

\bibitem[{{Kuerster} {et~al.}(1997){Kuerster}, {Schmitt}, {Cutispoto}, \&
  {Dennerl}}]{kuerster_1997}
{Kuerster}, M., {Schmitt}, J.~H.~M.~M., {Cutispoto}, G., \& {Dennerl}, K. 1997,
  \aap, 320, 831

\bibitem[{{Lalitha} {et~al.}(2013){Lalitha}, {Fuhrmeister}, {Wolter},
  {Schmitt}, {Engels}, \& {Wieringa}}]{lalitha_2013}
{Lalitha}, S., {Fuhrmeister}, B., {Wolter}, U., {et~al.} 2013, \aap, 560, A69

\bibitem[{{Pakull}(1981)}]{pakull_1981}
{Pakull}, M.~W. 1981, \aap, 104, 33

\bibitem[{{Ricker} {et~al.}(2015){Ricker}, {Winn}, {Vanderspek}, {Latham},
  {Bakos}, {Bean}, {Berta-Thompson}, {Brown}, {Buchhave}, {Butler}, {Butler},
  {Chaplin}, {Charbonneau}, {Christensen-Dalsgaard}, {Clampin}, {Deming},
  {Doty}, {De Lee}, {Dressing}, {Dunham}, {Endl}, {Fressin}, {Ge}, {Henning},
  {Holman}, {Howard}, {Ida}, {Jenkins}, {Jernigan}, {Johnson}, {Kaltenegger},
  {Kawai}, {Kjeldsen}, {Laughlin}, {Levine}, {Lin}, {Lissauer}, {MacQueen},
  {Marcy}, {McCullough}, {Morton}, {Narita}, {Paegert}, {Palle}, {Pepe},
  {Pepper}, {Quirrenbach}, {Rinehart}, {Sasselov}, {Sato}, {Seager},
  {Sozzetti}, {Stassun}, {Sullivan}, {Szentgyorgyi}, {Torres}, {Udry}, \&
  {Villasenor}}]{ricker_2015}
{Ricker}, G.~R., {Winn}, J.~N., {Vanderspek}, R., {et~al.} 2015, Journal of
  Astronomical Telescopes, Instruments, and Systems, 1, 014003

\bibitem[{{Schmitt} \& {Favata}(1999)}]{schmitt1999}
{Schmitt}, J.~H.~M.~M. \& {Favata}, F. 1999, \nat, 401, 44

\bibitem[{{Schmitt} {et~al.}(2019){Schmitt}, {Ioannidis}, {Robrade}, {Czesla},
  \& {Schneider}}]{schmitt2019}
{Schmitt}, J.~H.~M.~M., {Ioannidis}, P., {Robrade}, J., {Czesla}, S., \&
  {Schneider}, P.~C. 2019, \aap, 628, A79

\bibitem[{{Zechmeister} \& {K{\"u}rster}(2009)}]{Zechmeister2009}
{Zechmeister}, M. \& {K{\"u}rster}, M. 2009, \aap, 496, 577

\end{thebibliography}


\end{document}